\documentclass[sigconf, review=false, anonymous=false, nonacm=true]{acmart}
\usepackage{enumitem}
\AtBeginDocument{%
  \providecommand\BibTeX{{%
    \normalfont B\kern-0.5em{\scshape i\kern-0.25em b}\kern-0.8em\TeX}}}

\acmSubmissionID{303}

\begin{document}

\title{Chiplet Actuary: A Quantitative Cost Model and Multi-Chiplet Architecture Exploration}


\author{Yinxiao Feng, Kaisheng Ma}
\affiliation{%
  \institution{Institute for Interdisciplinary Information Sciences, Tsinghua University}
  \city{Beijing}
  \country{China}
}



\begin{abstract}
  Multi-chip integration is widely recognized as the extension of Moore's Law. Cost-saving is a frequently mentioned advantage, but previous works rarely present quantitative demonstrations on the cost superiority of multi-chip integration over monolithic SoC. In this paper, we build a quantitative cost model and put forward an analytical method for multi-chip systems based on three typical multi-chip integration technologies to analyze the cost benefits from yield improvement, chiplet and package reuse, and heterogeneity. We re-examine the actual cost of multi-chip systems from various perspectives and show how to reduce the total cost of the VLSI system through appropriate multi-chiplet architecture.
\end{abstract}


\maketitle

\section{Introduction}
Although Moore's Law has governed the semiconductor industry for over half a century, it is widely observed and recognized that Moore's Law is becoming harder to sustain. ``Integration of separately packaged smaller functions'' is considered the extension by Moore himself~\cite{Moore.2006} and the semiconductor industry.

The traditional VLSI system is implemented on a monolithic die, also known as system-on-chip (SoC). The growth of transistors on a single die is guaranteed by the steady growth of the process technology and the die area for the past few decades. However, as process technology improvement has slowed down and the chip area is approaching the limit of the lithographic reticle, transistor growth is going to stagnate~\cite{Loh.2021}\cite{Naffziger.2021}. Meanwhile, a large chip means more complex designs, and the poor yield results in even higher costs. Re-partitioning a monolithic SoC into several chiplets can improve the overall yield of dies, thereby reducing the cost.

Besides yield improvement, chiplet reuse is another characteristic of multi-chiplet architecture. In the traditional design flow, IP or module reuse is widely used; however, this approach still requires repeating system verification and chip physics design, which occupy a large part of the total non-recurring engineering (NRE) cost. Therefore, chiplet reuse, which saves the overhead of re-verifying systems and redesigning chip physics, can save more cost.

With the advent of many works about multi-chip, especially those products from the industry~\cite{Naffziger.2021}\cite{Xia.2021}, the economic effectiveness of multi-chiplet architecture has become a consensus. However, in practice, we find that the cost advantage of a multi-chip system is not easy to achieve due to the overhead of packaging and die-to-die (D2D) interface. Compared with SoC, the cost of multi-chip systems is much more difficult to evaluate at the early stage of VLSI system design. Without careful evaluation, adopting multi-chiplet architecture may lead to even higher costs. Previous works~\cite{Stow.2016}\cite{Stow.2017} focus on the manufacturing cost of dies and silicon interposers but neglect other significant costs such as substrates, D2D overhead, and NRE cost.

To better guide the VLSI system design and explain architecture challenges~\cite{Loh.2021} such as partitioning problem, we build a quantitative model \textit{Chiplet Actuary} for cost evaluation based on three typical multi-chip integration technologies. Based on this model, we discuss the total cost of different integration schemes from various perspectives. External data~\cite{Khan.2020}\cite{HIR}\cite{Li.2020}\cite{LLC_assembly}\cite{D5}\cite{ODSA} and in-house data are used to provide a relatively accurate final total cost. In summary, this paper makes the following major contributions to the VLSI system design:
\begin{itemize}[topsep=3pt, leftmargin=12pt, itemsep=2pt]
  \item We abstract monolithic SoC and multi-chip integration into different levels of concepts: module, chip, and package, by which we build a unified architecture.
  \item We present a quantitative cost model \textit{Chiplet Actuary} to estimate various components of the total system cost. To the best of our knowledge, this model is the first to introduce D2D overhead and NRE cost.
  \item Based on \textit{Chiplet Actuary}, we put forward an analytical method for decision-making on chiplet architecture problems: which integration scheme to use, how many chiplets to partition, whether to reuse packaging, how to leverage chiplet reusability, and how to exploit heterogeneity. Instructive insights are specified in Section \ref{summary}.
\end{itemize}

\section{Background}

\subsection{Multi-chip Integration}
\begin{figure}[b]
  \includegraphics[width=0.42\textwidth]{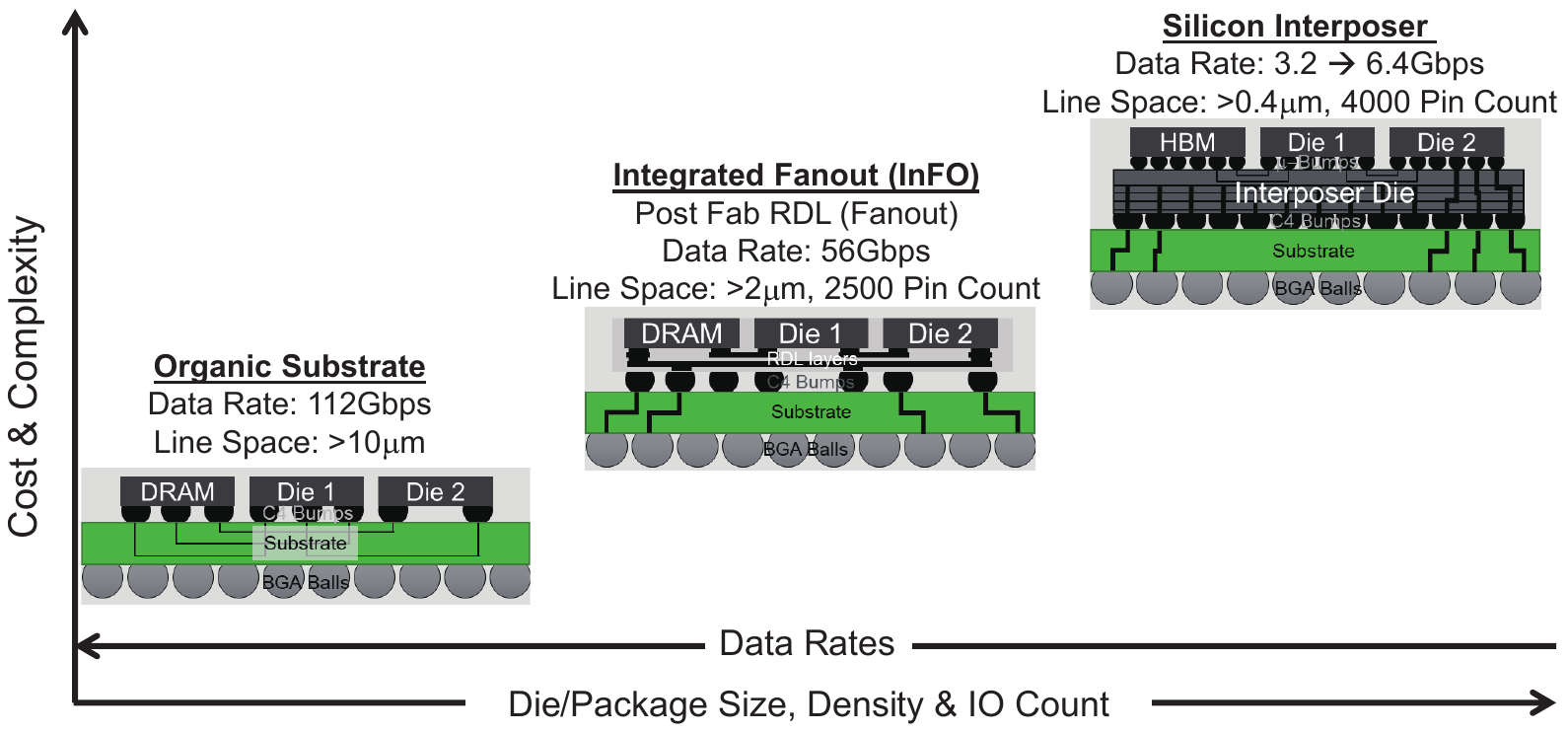}
  \caption{Different multi-chip integration technologies\protect~\cite{Synopsys} \label{synopsys}}
\end{figure}

Multi-chip integration is not an innovation but a technology developing over decades to make better VLSI systems. As shown in Figure \ref{synopsys}, the most widely used integration scheme is assembling different dies on a unifying substrate, also known as the typical multi-chip module (MCM) or system-in-package (SiP). Compared with MCM, integrated fan-out (InFO) technology is relatively more advanced. Developed from fan-out wafer-level packaging (FOWLP), InFO uses a redistribution layer (RDL) to offer smaller footprints and better electrical performance than the conventional substrate. According to the process sequence, InFO can be divided into chip-first and chip-last (or RDL-first).  In addition to 2D integration, silicon-interposer-based 2.5D integration, also called Chip-on-Wafer-on-Substrate (CoWoS) by TSMC, uses a relatively outdated chip to interconnect and integrate chiplets and memory dies. Though these three mainstream technologies are all used for multi-chip integration, they are different in package size, IO count, data rates, and cost. Therefore, chip designers are supposed to choose the right solution according to design objectives and cost constraints.

\subsection{Yield Model}
\begin{figure}[htbp]
  \setlength{\abovecaptionskip}{5pt}
  \includegraphics[width=0.39\textwidth]{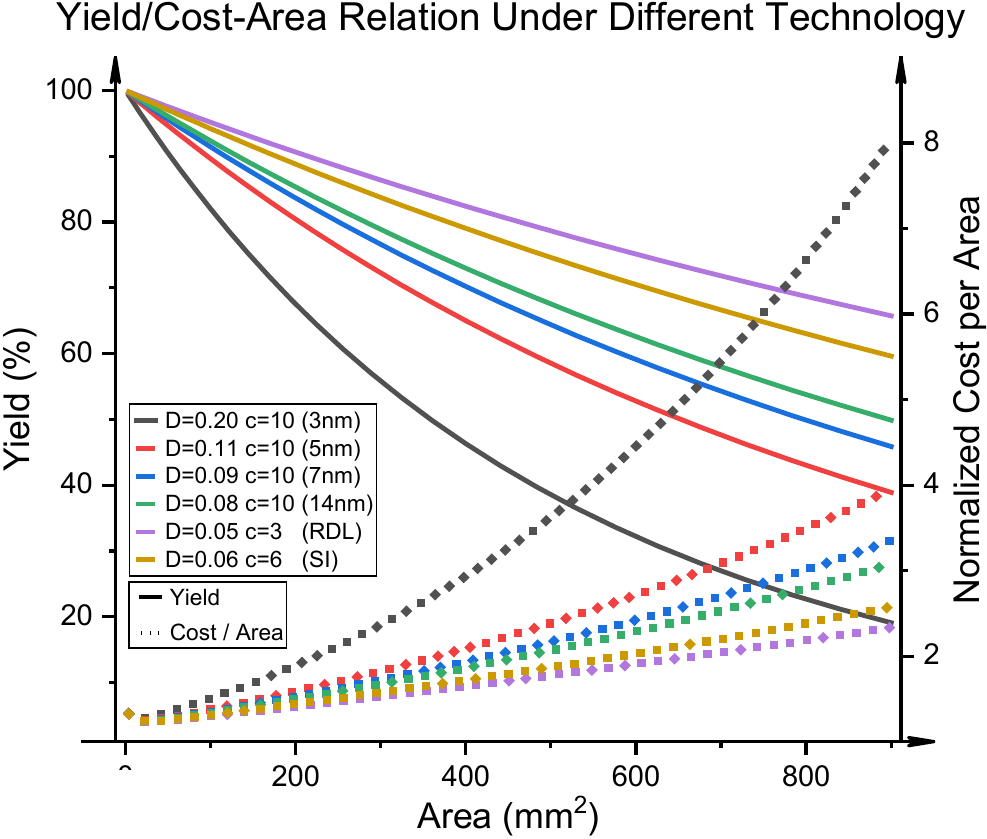}
  \caption{Yield/Cost-Area relation of different technologies \label{Yield}}
\end{figure}

One of the core components of the cost model is the yield model, which has been an important topic since the advent of the integrated circuit industry. For predicting yields of dies, Poisson, Negative Binomial, and other models from the industry are used to provide a more accurate result. Among these models, Seed's model and the Negative Binomial model are the most widely used in the same form of~\cite{Cunningham.1990}
\begin{equation}
  {\rm Yield_{~die}}=\left(1 + \frac{DS}{c}\right)^{-c} ,
\end{equation}
where $D$ is the defeat density, $S$ is the die area, and $c$ is the cluster parameter in the Negative Binomial model or the number of critical levels in Seed's model. We have followed this model and used more realistic parameters. Figure \ref{Yield} shows the yield-area and the cost-area relations of different technologies under this model. All costs are normalized to the cost per area of the raw wafer.

The traditional SoC is manufactured in a serial production line, so the overall yield is estimated by continuous multiplication
\begin{equation}
  Y_{\rm ~overall} = Y_{\rm ~wafer} \times Y_{\rm ~die} \times Y_{\rm ~packaging} \times Y_{\rm ~test} .
\end{equation}
However, for the multi-chip system, yield cannot be estimated by simple multiplication because of the more complex manufacturing flow.

\subsection{NRE and RE Cost}
The total cost of VLSI systems can be roughly divided into two kinds: non-recurring engineering (NRE) cost and recurring engineering (RE) cost. NRE cost refers to the one-time cost of designing a VLSI system, including software, IP licensing, module/chip/package design, verification, masks, etc. RE cost refers to the fabrication costs for massive production, including wafers, packaging, test, etc.

For one VLSI system, its final engineering cost consists of the RE and the amortized NRE cost. Amortization is mainly related to the proportion of quantity. The basic concept is that if the production quantity is small, the NRE cost is dominant; otherwise, the NRE cost is negligible if the quantity is large enough.

\section{Chiplet Actuary Model}

\subsection{High Level Abstraction}
Our model is implemented for comparing the RE and NRE cost between monolithic SoC and multi-chip integration. As the problem is so complex, we use some necessary assumptions to ignore non-primary factors:
\begin{itemize}[topsep=3pt, leftmargin=15pt, itemsep=2pt]
  \item All chiplets under the same process node share the same die-to-die (D2D) interface with different channel numbers;
  \item Performance and power are not considered in this model;
  \item Different parts of the NRE cost are independent so that they can be estimated separately.
\end{itemize}

Besides the above assumptions, many other approximations are also used in the model. More details can be referred to in our open-source code of the model\footnote{Repository URL: https://github.com/Yinxiao-Feng/DAC2022.git}.

\begin{figure}[htbp]
  \includegraphics[width=0.41\textwidth]{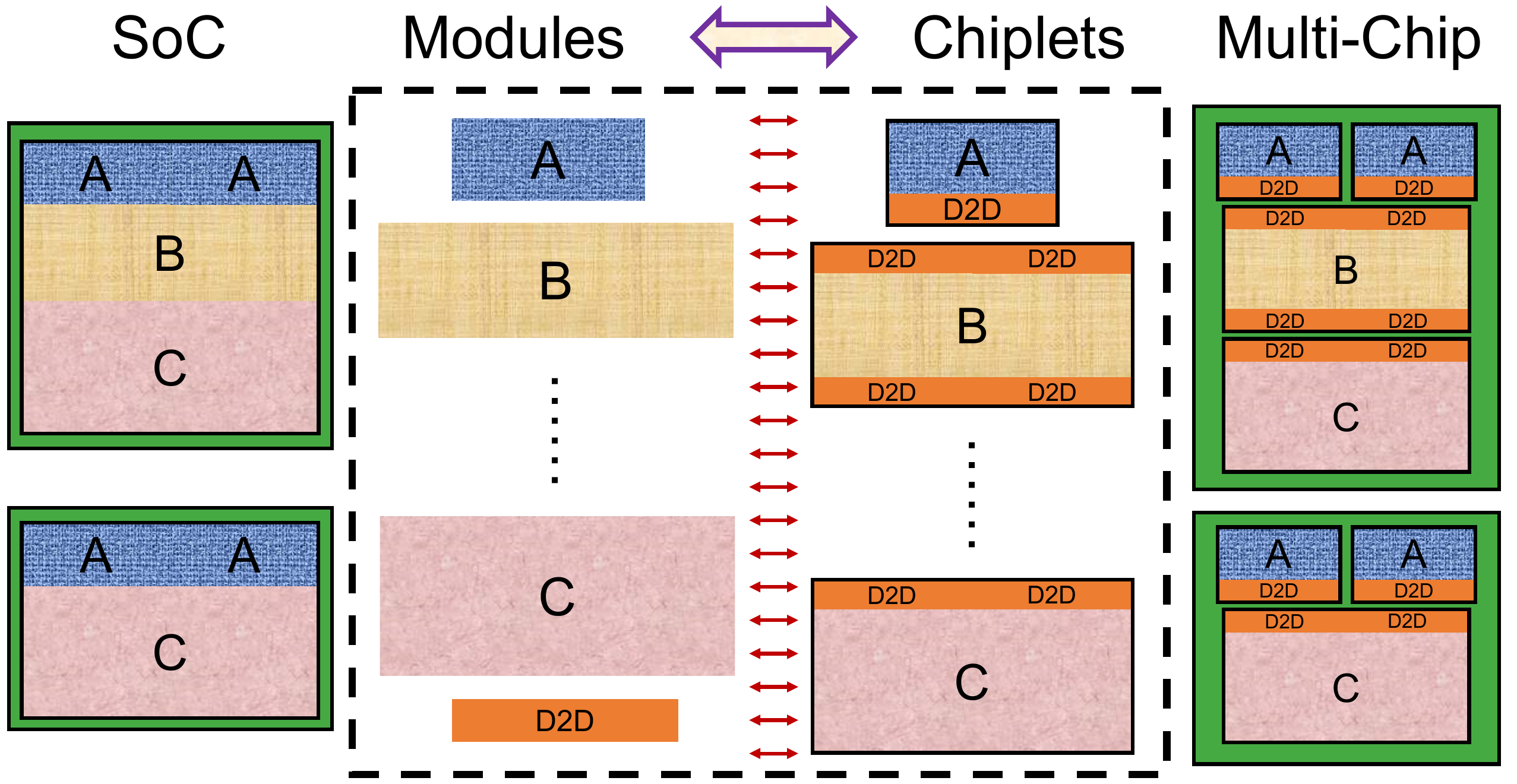}
  \caption{High-level cost model diagram \label{HL}}
\end{figure}

As shown in Figure \ref{HL}, module, chip and package are the three main concepts involved in our model. A group of systems is built from a group of modules. Each module corresponds to a chiplet. Each system can be SoC formed directly from modules or multi-chip integration formed from chiplets. The relation can be described as follows:
\begin{equation}
  \begin{aligned}
    m_i \in       & \{m_1, m_2, ... , m_{D2D}\} = M                         \\
    c_i =         & ~{\rm Chip}(\{m_i, m_{D2D}\}) \in C                     \\
    {\rm SoC}_j = & ~{\rm Package}({\rm Chip}(\{m_{k_1}, m_{k_2}, \dots\})) \\
    {\rm MCM}_j = & ~{\rm Package}(\{c_{k_1}, c_{k_2}, \dots\}) ,
  \end{aligned}
\end{equation}
where $m$ and $c$ are module and chiplet, $\rm Package(\cdot)$ and $\rm Chip(\cdot)$ are methods forming system from chips and forming chip from modules. Different from the general concept of the module, our module refers to an indivisible group of functional units. D2D interface is a particular module with which each module makes up a chiplet. D2D interfaces under different process nodes are regarded as diverse modules.

\subsection{RE Cost Model}
The RE cost in our model consists of five parts: 1) cost of raw chips, 2) cost of chip defects, 3) cost of raw packages, 4) cost of package defects, 5) cost of wasted known good dies (KGDs) resulting from packaging defects. Other costs such as bumping, wafer sort, and package test are also included but not itemized separately because they are not so significant~\cite{Stow.2016}\cite{Stow.2017}.

On the basis of previous works~\cite{Stow.2016}\cite{Stow.2017}, we make several improvements. The first is the consideration of D2D interface overhead. For any multi-chip system, especially those with high interconnection bandwidth, the D2D interface occupies a considerable portion of the area~\cite{ODSA}. In our model, we regard D2D interface as a particular module shared by all chiplets. It takes a certain percentage of the chip area depending on different technologies and architectures.

Then, more multi-chip integration models are included. MCM is similar to SoC that flips chips directly on a unified organic substrate. The difference is that the MCM needs additional substrate layers for interconnection, so MCM has a growth factor on substrate RE cost. As for InFO and 2.5D, the interposer cost is calculated similarly with the die cost, and the bump cost and bounding yield are counted twice on the chip side and the substrate side. The total cost resulting from packaging is
\begin{equation}
  \begin{aligned}
    \rm
    Cost_{~packaging} = & \rm  ~Cost_{~Raw~Package}                                                             \\
    +                   & {\rm  ~Cost_{~interposer}} \times \left(\frac{1}{y_1 \times y_2^n\times y_3}-1\right) \\
    +                   & {\rm  ~Cost_{~substrate}} \times \left(\frac{1}{y_3}-1\right)                         \\
    +                   & {\rm  ~Cost_{~KGD}} \times \left(\frac{1}{y_2^n \times y_3}-1\right) ,
  \end{aligned}
\end{equation}
where $y_1$ is the yield of the interposer, $y_2$ is the bonding yield of chips, $y_3$ is the bonding yield of the interposer. The difference between chip-first and chip-last is also viewed. As shown in the equations
\begin{equation}
  \begin{aligned}
    \rm
    Cost_{~chip-first} & \rm = \frac{\sum{\frac{C_{chip}}{Y_{chip}}} + C_{package}}{Y_{package}}                                                        \\
    \rm
    Cost_{~chip-last}  & = \frac{\rm \frac{C_{package}}{Y_{package}}+\sum{\left(\frac{C_{chip}}{Y_{chip}}+C_{bond}\right)}}{\text{Y}_{\rm bonding}^n} ,
  \end{aligned}
\end{equation}
though chip-first packaging flow is simpler, the poor yield of packaging would result in a huge waste on KGDs. Therefore, chip-last packaging is the priority selection for multi-chip systems, and our experiments below are based on it.

We break down various components of the total RE cost to better analyze the reason behind it. As we find that the cost of wasted KGDs resulting from packaging takes a significant proportion of the total cost, especially when the die cost is high and the packaging yield is poor, this part of the cost is counted separately.

\subsection{NRE Cost Model}
NRE cost is rarely discussed quantitatively in previous works because it depends on the particular circumstances of each design team. As it is so essential, in any case, we need to build a model to guide us in designing VLSI systems.

We use the area as the unified measure. In our model, the NRE cost consists of three parts: 1) cost for designing modules, 2) cost for designing chips, 3) cost for designing package. For any chip $c$, the NRE cost can be estimated by the equation
\begin{equation}
  {\rm Cost}  =  K_cS_c + \sum_{m_i \in c}{K_{m_i}S_{m_i} + C} ,
\end{equation}
where $S_c$ is the area of the chip and $S_{m_i}$ is the area of module $i$. $K_c$ and $K_m$ are the factors associated with design complexity and design capability. $K_c$ is determined by NRE costs related to the chip area, such as system verification and chip physics design; $K_m$ is determined by NRE costs related to the module area such as module design and block verification; $C$ is the fixed NRE costs for each chip independent of area, such as IP licensing and full masks. The NRE model can reflect the difference between module-reuse-based SoC and chiplet-reuse-based multi-chip integration. For a group of systems $J$ built by monolithic SoC, the total NRE cost can be expressed as

\begin{equation}
  \begin{aligned}
    {\rm Cost} & =  \sum_{j \in J}{(K_{c_j}S_{c_j} + C_j + K_{p_j}S_{p_j} + C_{p_j})} \\
               & + \sum_{m_i \in M}K_{m_i}S_{m_i} ,
  \end{aligned}
\end{equation}
where $K_{p_j}$ is the cost factor of the system $j$ related to the integration technology, $S_p$ is the package area and $Cp$ is the fixed NRE cost for each package independent of area. The same module needs to be designed only once, but every chip needs to be individually designed. If we build these systems by multi-chip integration, the total NRE cost changes into

\begin{equation}
  \begin{aligned}
    {\rm Cost} & = \sum_{j \in J}{(K_{p_j}S_{p_j} + C_{p_j})}              \\
               & + \sum_{c_i \in C}(K_{c_i}S_{c_i} + K_{m_i}S_{m_i} + C_i) \\
               & + \sum_{n}{C_{
          {\rm D2D}_n}} ,
  \end{aligned}
\end{equation}
where $C_{{\rm D2D}_n}$ is the NRE cost for designing D2D interface under process node $n$. It is obvious that multi-chip integration benefits not only from module reuse but also from chip reuse.

\begin{figure*}[tb]
  \setlength{\textfloatsep}{0pt}
  \includegraphics[width=0.98\textwidth]{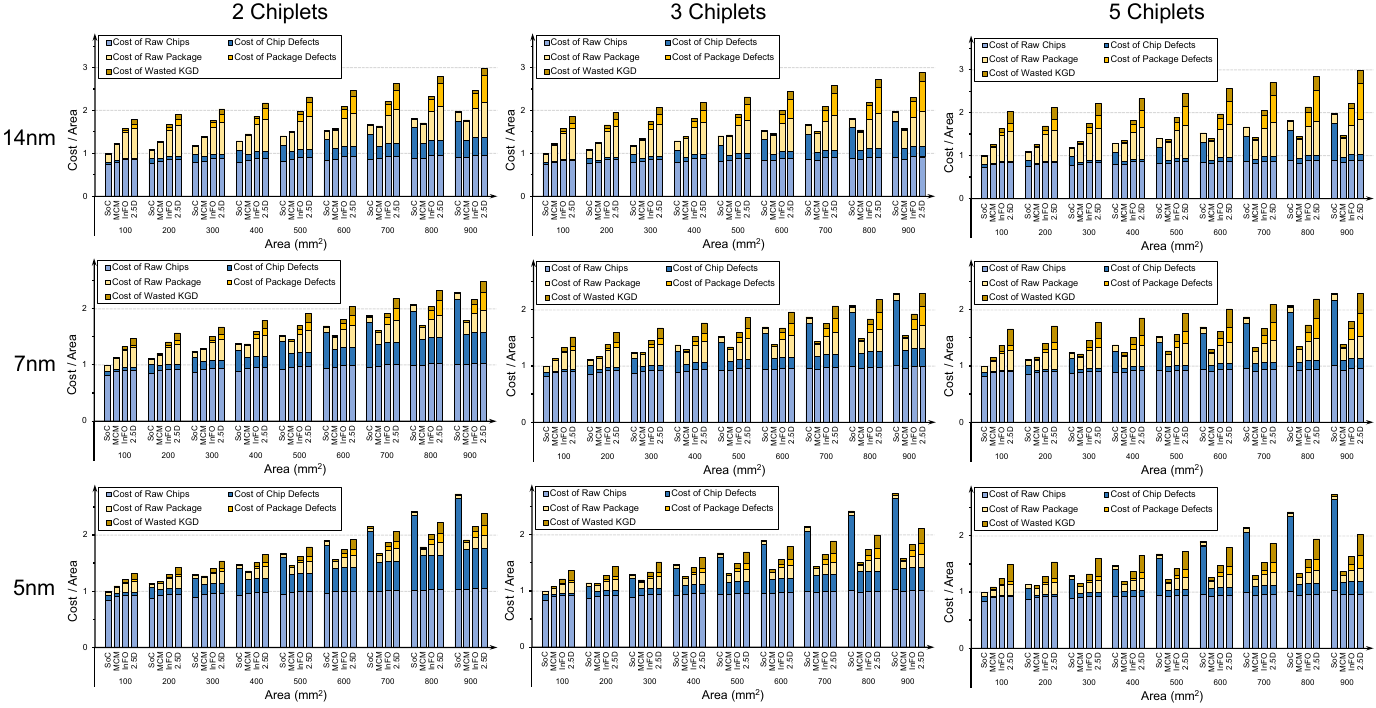}
  \caption{Normalized RE cost comparison among different integrations under different technologies \label{RE_Cost}}
\end{figure*}

\section{Model Validation and Discussion}
Data used in the experiments is from commercial databases~\cite{LLC_assembly}, public information~\cite{Khan.2020}\cite{HIR}\cite{Li.2020}\cite{D5}\cite{ODSA}, and the in-house. The experiment results are convincing under these situations, but applying the model to other cases makes it necessary to include the latest relevant data as the parameters of the model.

\subsection{Validation and Comparison of RE cost}
\label{RE}

\begin{figure}[tb]
  \includegraphics[width=0.37\textwidth]{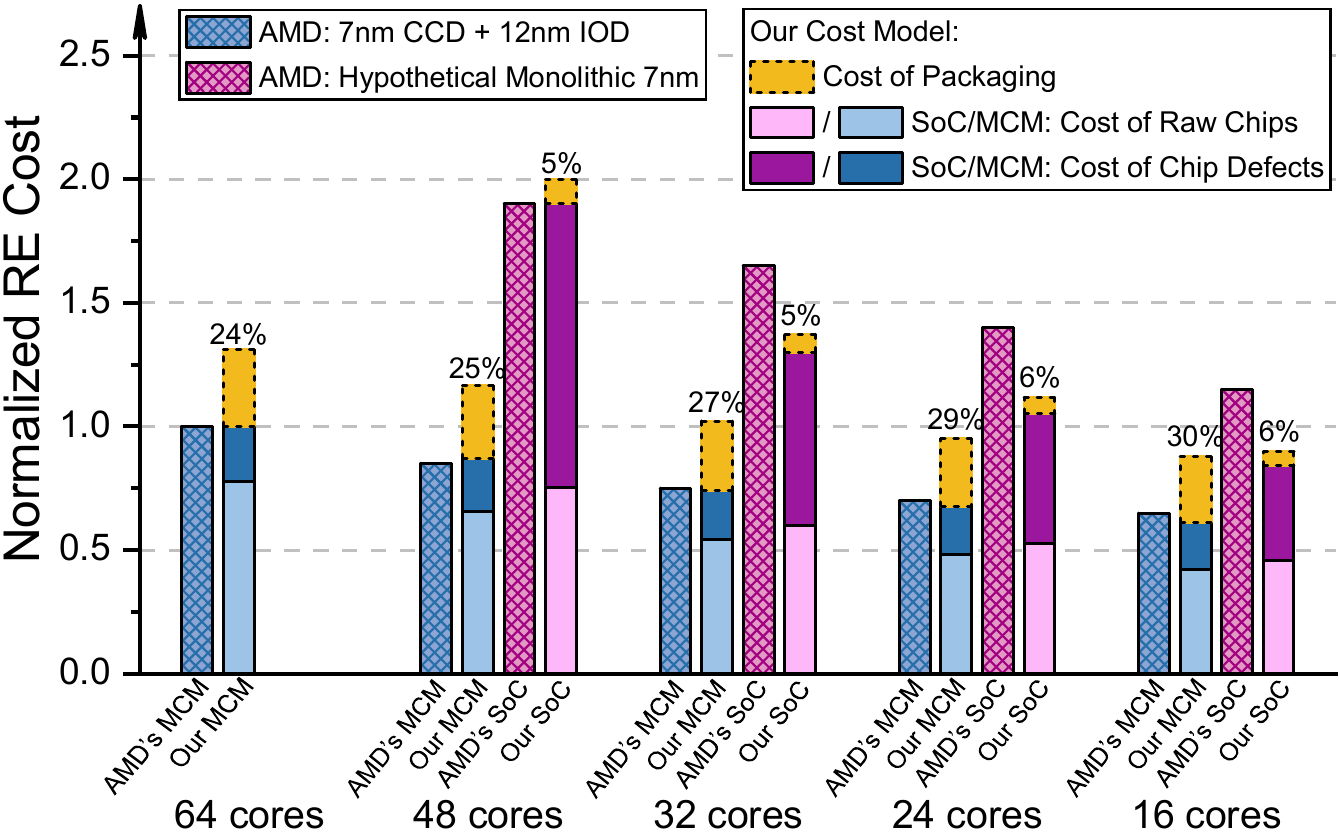}
  \caption{Normalized RE cost comparison for AMD's chiplet architecture.\label{AMD} \protect \footnotemark }
\end{figure}

We validate our model on public works. AMD comes up with the well-known chiplet architecture~\cite{Naffziger.2021}. As Figure \ref{AMD} shows, AMD claims that their chiplet-based products have a considerable cost advantage over monolithic SoC. We validate our model on AMD's design based on external and in-house data. Considering that the TSMC 7nm and GF 12nm process has just been massive-produced when the Zen3 project is initiated, relative high defect density parameters (0.13 for 7nm, 0.12 for 12nm, speculate based on public data~\cite{D5}) are used. The comparison shows die costs result similar with AMD. Multi-chip integration can save up to 50\% of the die cost. However, AMD covers up their additional costs for reintegration. When taking packaging overhead into account, the advantages of multi-chip are reduced. Especially for the 16 core system, the packaging cost accounts for 30\%. As the yield of 7nm technology improves in recent years, the advantage is further smaller.

\footnotetext{The cost of packaging is the sum of the cost of raw package, the cost of package defects, and the cost of wasted KGDs.}

Based on recent data, more explorations on RE cost under various integrations and technologies are studied. We divide a monolithic chip into different numbers of chiplets and then assembly them by various integration methods. Referring to EPYC~\cite{Naffziger.2021}, 10\% of the D2D interface overhead is assumed, and no reuse is utilized. All costs are normalized to the 100 $mm^2$ area SoC.

As Figure \ref{RE_Cost} shows, there are significant advantages for advanced technology (5nm) because the cost resulting from die defects accounts for more than 50\% of the total manufacturing cost of the monolithic SoC at 800 $mm^2$ area. As for mature technology (14nm), though there are also up to 35\% cost-savings from yield improvement, the cost advantage of multi-chip is not that significant because of the D2D and packaging overhead (>25\% for MCM, >50\% for 2.5D). For any technology node, the benefits increase with the increase of area, and the turning point for advanced technology comes earlier than the mature technology. As InFO and 2.5D based multi-chip integration consist of a large monolithic interposer, they also suffer from the poor yield of the complex packaging process; moreover, bonding defects lead to waste of KGDs, so the cost of packaging (50\% at 7nm, 900$mm^2$, 2.5D) is comparable with the chip cost. Therefore, advanced packaging technologies are only cost-effective under advanced process technology.

Another important insight is about granularity. The cost benefits from smaller chiplet granularity have a marginal utility. With the increase of chiplets quantity (3$\rightarrow$5), the cost-saving of die defects is more negligible (<10\% at 5nm, 800$mm^2$, MCM), and the overhead is higher.

\subsection{Total Cost Comparison of Single System}
\label{single_system}

Though RE cost is a major cost to be considered, the NRE cost is often the determinant, especially for systems without huge production guarantees. Take a system of 800$mm^2$ module area as an example. We implement the system by monolithic SoC and two chiplets MCM separately. D2D overhead is also assumed at 10\%. NRE cost is amortized to each system depending on the number of modules and chips included. All cost is normalized to the RE cost of SoC.

\begin{figure}[tb]
  \includegraphics[width=0.47\textwidth]{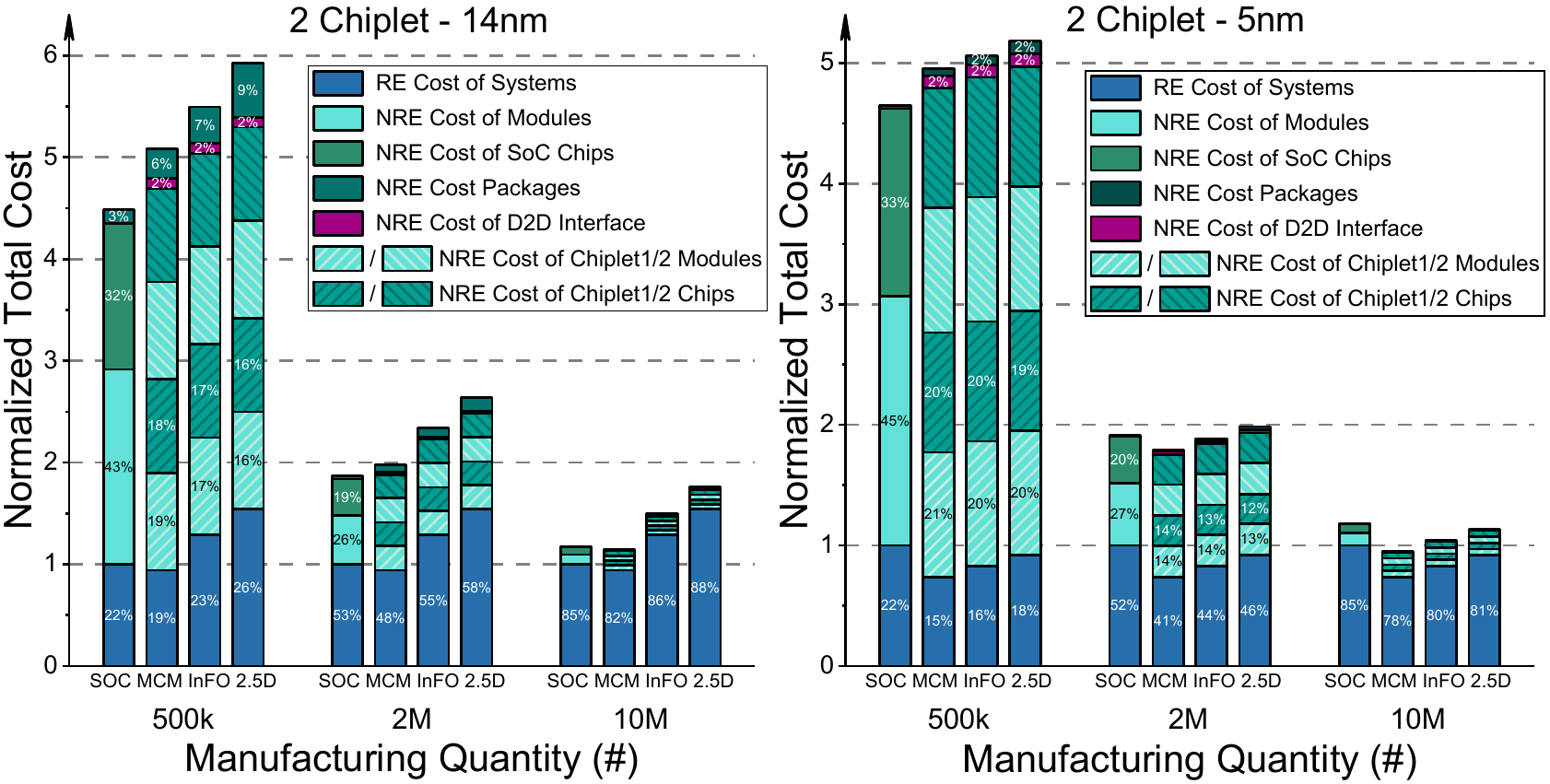}
  \caption{Normalized total cost structure of single system}
  \label{single}
\end{figure}

As shown in Figure \ref{single}, because of the large total module area, the NRE overhead of D2D interface and packaging is no more than 2\% and 9\% (2.5D), and the total NRE cost for designing modules also remains the same. However, for each chiplet, there is a high fixed NRE cost, such as masks, hence multi-chip leads to very high NRE costs (36\% at 500k quantity) for designing and manufacturing chips. For 5nm systems, when the quantity reaches two million, multi-chip architecture starts to pay back. As for smaller systems, the turning point of production quantity is further higher. So, monolithic SoC is often a better choice for a single system unless the area or the production quantity is large enough.

\section{Chiplet Reuse Scheme Exploration}
\label{sec_reuse}

There are several common ways of chiplet reuse in the industry such as EPYC\cite{Naffziger.2021} and LEGO~\cite{Xia.2021}. In this section, we will show how these architectures achieve cost benefits. From the explorations, we can appropriately adopt multi-chiplet architectures.

\begin{figure}[h]
  \includegraphics[width=0.45\textwidth]{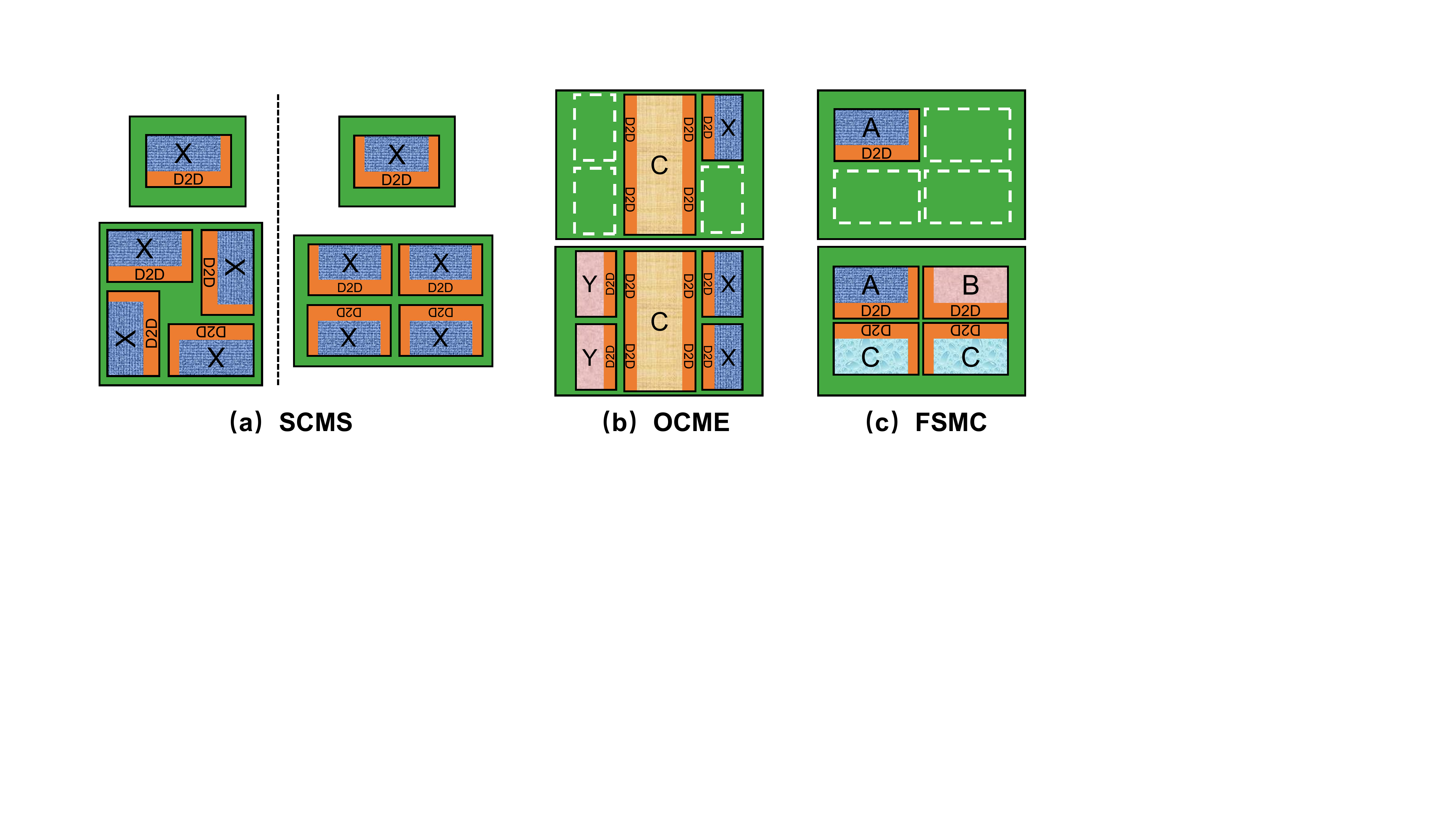}
  \caption{Different reuse schemes \label{reuse_pattern}}
\end{figure}

\subsection{Single Chiplet Multiple Systems (SCMS)}
\label{sec_SCMS}
As shown in Figure \ref{reuse_pattern}(a), SCMS is a multi-chip architecture that uses a single kind of chiplet to build several systems\footnote{Symmetrical placement requires a symmetrical chiplet; otherwise, two mirrored chiplets are necessary.}. We take a 7nm chiplet with 200$mm^2$ module area as an example. Three systems containing 1, 2, and 4 chiplets are built based on MCM and 2.5D, and the production quantity for each system is assumed at 500,000. Two conditions with or without package reuse are also considered. All costs are normalized to the RE cost of the 4X MCM system.

\begin{figure}[tb]
  \includegraphics[width=0.4\textwidth]{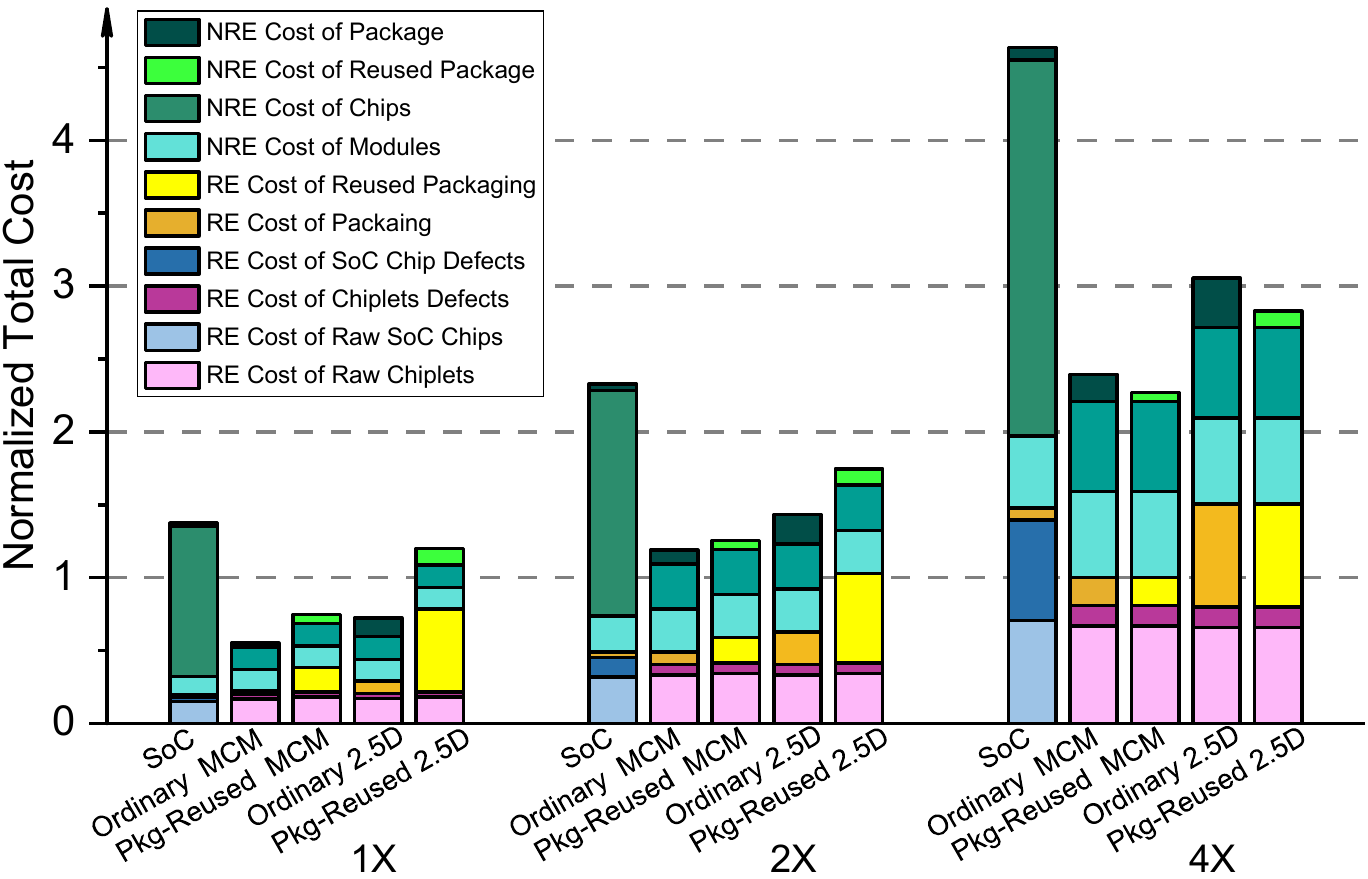}
  \caption{Normalized total cost of SCMS reuse scheme\label{SCMS}}
\end{figure}

As Figure \ref{SCMS} shows, due to chiplet reuse, there is vast chip NRE cost-saving (nearly three quarters for 4X system) compared with monolithic SoC. The advantage of the SCMS reuse scheme is that only one chiplet is needed, so it comes into effect instantly without making multiple chips. This architecture is suitable for one production line with different grades. The disadvantage is that D2D interconnections lead to significant overhead, and as there is only one kind of chiplet, there is no possibility for heterogeneous technology.

If the package is reused among these three systems, for the largest 4X system, the NRE cost of the package will be reduced by two-thirds. However, for the smallest 1X system, the total cost will increase more than 20\%. Package reuse saves amortized NRE cost of package for larger systems but wastes RE cost for smaller systems. Therefore, whether using package reuse depends on which accounts for a more significant proportion.

For advanced packaging such as 2.5D, if the 4x interposer is reused in the 1x system, packaging cost more than 50\%. Therefore, package reuse is uneconomic for high-cost 2.5D integrations, but 2.5D can still benefit from chiplet reuse.

\subsection{One Center Multiple Extensions (OCME)}
\label{sec_OCME}
As Figure \ref{reuse_pattern}(b) shows, in the OCME architecture, there are a reused die (C) in the center and various extension chips with the same footprint placed around. We take a 7nm 4-160$mm^2$-sockets system as an example. Two different extension dies \{X, Y\} are used to build four different systems, and the production quantity for each system is assumed at 500,000. Both with and without package reuse are taken into account, and all cost is normalized to the RE cost of the largest MCM system. We also perform experiments on the possibility that the center die can be designed under relatively outdated process technology (14nm).

\begin{figure}[b]
  \includegraphics[width=0.41\textwidth]{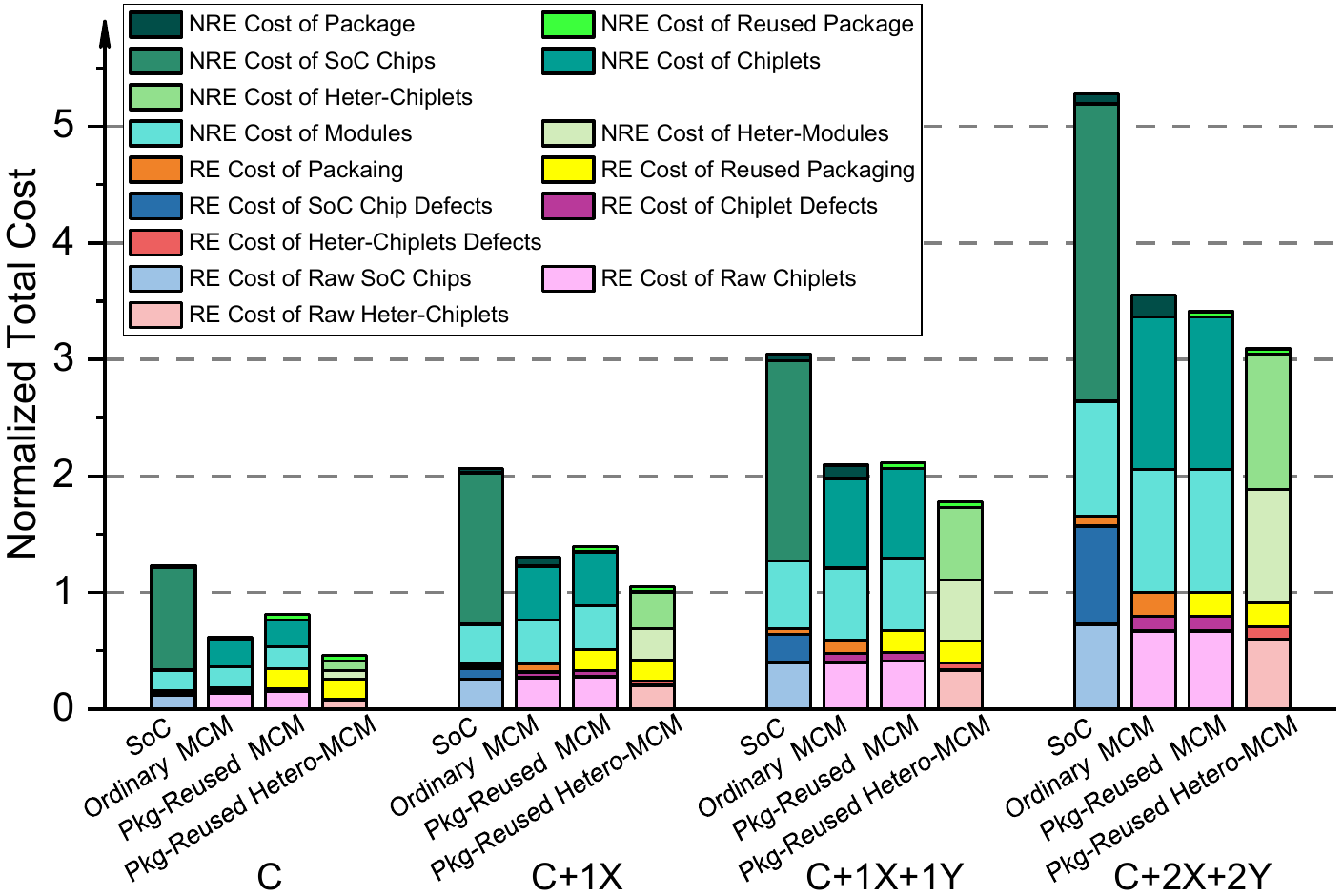}
  \caption{Normalized total cost of OCME reuse scheme}
  \label{OCME}
\end{figure}

Figure \ref{OCME} shows the amortized total cost of SoC, ordinary MCM, package reused MCM, and package reused heterogeneous MCM. The reuse benefit is not as evident (NRE cost-saving < 50\%) as the SCMS scheme because three chiplets are used, and the average reuse is less. Therefore, the OCME scheme needs more systems to come into effect.

The advantage of the OCME reuse scheme is the possibility of heterogeneity. With heterogeneous integration, shown in Figure~\ref{OCME}, the total costs are further reduced by more than 10\%. Especially for the single C system, there is almost half the cost-saving. For systems that share a large area of modules that do not benefit from advanced process technology nodes, adopting the OCME scheme is more cost-effective.

\vspace{-2pt}
\subsection{A few Sockets Multiple Collocations (FSMC)}
\label{sec_FSMC}
Besides the two schemes above, a package with several chip sockets can hold more systems. As shown in Figure \ref{reuse_pattern}(c), assume there are $n$ different chiplets with the same footprint, and the package has $k$ sockets. It follows that up to $\sum_{i=1}^{k}{(C_{n+i-1}^{i})}$ different systems can be built. It only takes six chiplets and one 4-sockets package to build a maximum of up to 119 diverse systems. We ideally assume that all of these reuse possibilities are utilized, and each system has a production quantity of 500,000. Five different situations from low to high reuse times are compared by average normalized cost.

\begin{figure}[tb]
  \includegraphics[width=0.34\textwidth]{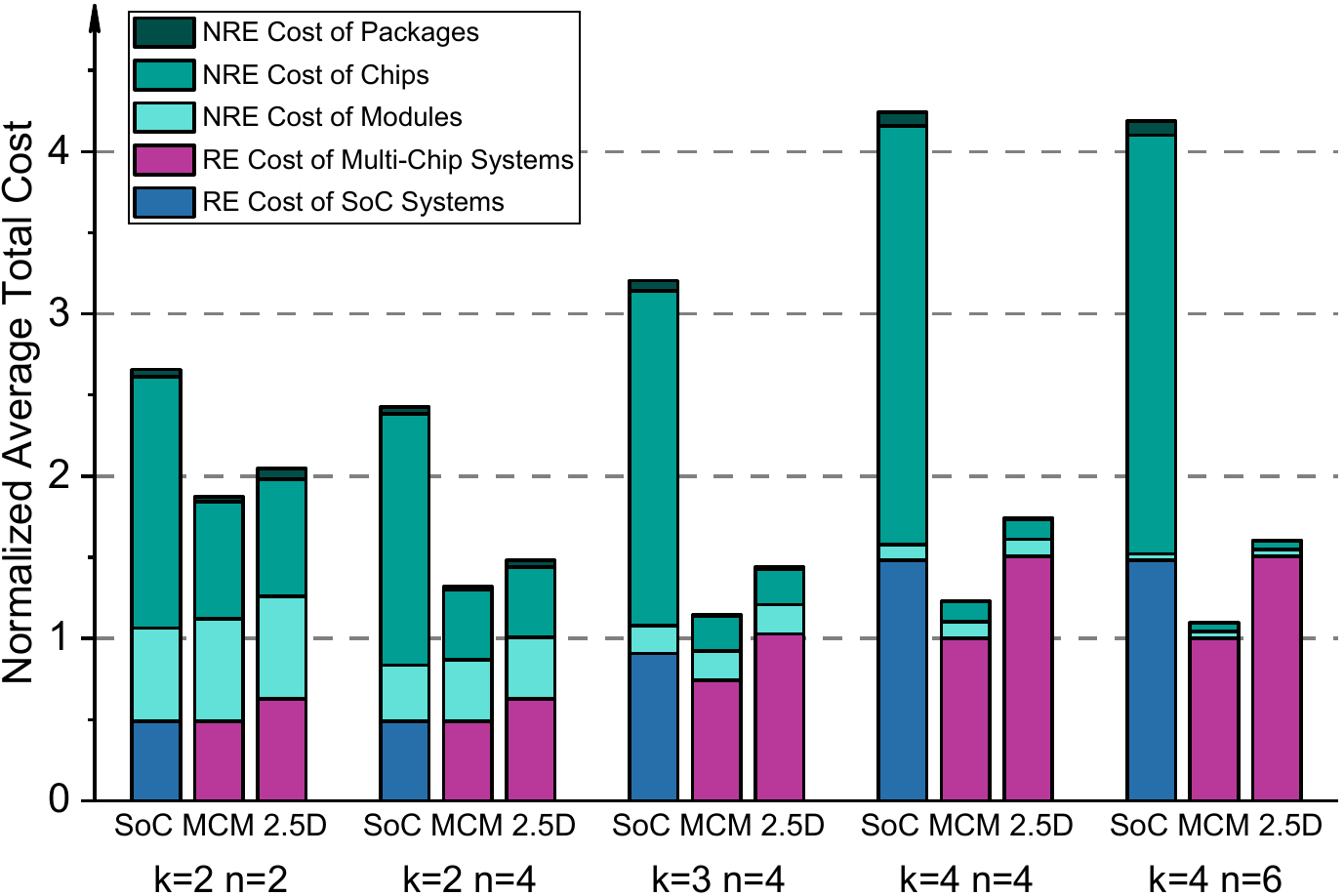}
  \caption{Normalized total cost of FSMC reuse scheme}
  \label{FSMC}
\end{figure}

As shown in Figure \ref{FSMC}, the more chiplets are reused, the more benefits from NRE cost amortization. When the reusability is taken full advantage of, the amortized NRE cost is small enough to be ignored. At this point, the huge cost-saving potential for multi-chip architecture is revealed. Cost advantage is achieved not only from the RE cost-saving but also the NRE cost-saving.

The models above are the most idealized result because not every chiplet can be reused many times, and not every system has actual demand. After all, few systems can have billions of production quantities, so the space for NRE cost amortization always exists. Given that most chip design teams have limited design capabilities and production quantities, it may be more economical to build systems by a few excellent chiplets from different specialized teams.

\vspace{-2pt}
\section{Summary}
\label{summary}
Multi-chip architecture has become a future trend. However, from our point of view, the benefit of multi-chip architecture is not unconditional but depends on many complicated factors. To help chip architects make better decisions on multi-chip architectures, we build a quantitative model for cost comparison among different alternatives. Our model allows designers to validate the cost at the early stage. We have also shown how multi-chip architecture can actually benefit from yield improvement, chip and package reuse, and heterogeneity. The takeaways of this paper are summarized as follows:
\begin{itemize}[topsep=3pt, leftmargin=10pt, itemsep=2pt]
  \item Multi-chip architecture begins to pay off when the cost of die defects exceeds the total cost resulting from packaging; The closer to the \textit{Moore Limit} (the largest area at the most advanced technology) the system is, the higher cost-benefit from multi-chip architecture is. RE cost benefits from smaller chiplet granularity have marginal utility, so splitting a single system into two or three chiplets is usually sufficient. (Section \ref{RE})
  \item For a single system, monolithic SoC is a better choice unless the production quantity is large enough to amortize the NRE overhead of multiple chiplets. (Section \ref{single_system})
  \item Whether to reuse packaging depends on whether the RE or the amortized NRE cost is dominant.  (Section \ref{sec_SCMS}, \ref{sec_OCME})
  \item For systems of multiple grades, the SCMS scheme brings significant cost advantages; For systems that share a large area of ``unscalable'' modules, adopting the OCME scheme is more cost-effective; the FSMC scheme provides maximum reuse possibilities.  (Section \ref{sec_reuse})
  \item The basic principle is building more systems by fewer chiplets, and the cost benefits of chiplet reuse are more evident for finely segmented demands.   (Section \ref{sec_FSMC})
  \item Despite all the benefits, unfortunately, Moore's Law has not been fundamentally extended. For ultra-high performance systems which are close to the \textit{Moore Limit}, the interconnection requirements are too high to be supported by the organic substrate, so advanced packaging technologies such as InFO and 2.5D are necessary. However, with a monolithic interposer, advanced packaging technologies still suffer from poor yield and area limit.
\end{itemize}

\section{Acknowledgement}
The authors would like to thank ANSYS Inc. for useful EDA software, Polar Tech Inc. for latest in-house data and other colleagues for their strong support.

\bibliography{reference}

\end{document}